# Quality control of eggs using multivariate analysis of micro-Raman spectroscopy

Maryam Davari [1], Maryam Bahreini[1*], Zahra Sabzevari[1]

[1] School of physics, Iran university of science and technology, Tehran, Iran

*Corresponding author email: m_bahreini@iust.ac.ir

**Abstract**

Considering the pivotal role of eggs in the food industry and their nutritional significance, this study employed micro-Raman spectroscopy of eggs, examining both shells and yolks to assess the quality and freshness of eggs. To investigate temperature and time effects, Raman spectra were collected at different temperature and time intervals, potentially indicating Raman peaks reduction due to Maillard reaction and oxidation of proteins and lipids and carotenoid depletion, respectively. By calculating the ratio of Raman peaks, lipids, fatty acids, and choline methyl were introduced as biomarkers of temperature and time. Notable correlations were identified between Raman peaks and egg quality coefficients, including egg coefficient and peak 1002 cm$^{-1}$ (protein), total weight and 1301 cm$^{-1}$ (Lipids), yolk weight and 2934 and 3057 cm$^{-1}$, total weight with peak 710 cm$^{-1}$, and egg shape index and peak 3057 cm$^{-1}$. Analysis of eggshells at different time intervals revealed Raman peak reduction during time, demonstrating Raman's effectiveness in assessing egg quality from its shell. Using PLS-DA method, the classification of eggs at different temperature and storage time using egg yolk Raman spectra was performed with 80% accuracy, predominantly influenced by carotenoid peaks, showing Raman a practical, and non-destructive method for egg quality and freshness control.

## 1. Introduction

Today, the egg industry holds a significant position in the global food industry (Kopec *et al.*, 2022), eggs being recognized as a 'whole nutrition food' and a high-quality source of animal protein (Zang *et al.*, 2023). Rich in essential lipids, proteins, minerals, and trace elements, eggs also serve as valuable sources of vitamin A, iron, vitamin B12, riboflavin, and choline (Réhault-Godbert, Guyot and Nys, 2019)(Roseland *et al.*, 2020). Notably, eggs are acknowledged as a substantial source of vitamin D, including D3 and 25(OH)D3 (Roseland *et al.*, 2020).

The nutritional composition of the egg yolk, is characterized by approximately 50% water, 30% lipids, and 16% proteins, with minor amounts of carbohydrates and minerals (Wang *et al.*, 2020). The egg yolk also contains various antioxidants, such as carotenoids (Kopec *et al.*, 2022), which play a crucial role in biological functions, acting as antioxidants and contributing to the reduction of cancer risk, coronary heart disease, stroke, type 2 diabetes mellitus, and asthma in adults and children (Carvalho *et al.*, 2019a; Udensi *et al.*, 2022).

Given the importance of eggs in the human diet, ensuring their quality is paramount. Over time, stored eggs experience a decline in quality, leading to the degradation of nutritional and functional components, the development of undesirable flavors, and an increased presence of spoilage and pathogenic microorganisms, such as Salmonella (Wu *et al.*, 2023). Egg quality encompasses factors such as size, shell color, shell quality, egg shape, internal quality, and inclusions like blood and meat spots (Kemps *et al.*, 2006). internal quality of the egg, is shown by the Haugh unit, yolk coefficient, yolk color, shell weight, strength, thickness, pH of albumen, egg shape index, and yolk index (Karoui *et al.*, 2006; Liu *et al.*, 2007; Abdanan Mehdizadeh *et al.*, 2014; Zita, Jeníková and Härtlová, 2018).

Molecular vibration spectroscopy is occupied in various fields currently, including medicine, agriculture, and food owing to its fast detection speed and simple sample treatment (Xu *et al.*, 2024). Nowadays spectroscopic techniques have become indispensable and important technological means for food quality and safety inspection (Khoshroo *et al.*, 2015). Raman spectroscopy, among these spectroscopic techniques, has acceptable properties, such as rapidness, robustness, and non-destructiveness likewise the ability to provide distinct fingerprint information (Zhu *et al.*, 2023a). Different molecules in a couple number of samples can be detected without the need for any reagent by Raman spectroscopy (Bahreini *et al.*, 2019)(Bahreini, 2018). Also, the identification of biochemical groups such as amide bands, methyl groups, and ringed structures on subcellular structures such as nucleic acids and proteins is one of the superiorities of Raman spectroscopy (Maryam Bahreini, 2015). Therefore, Raman spectroscopy has a significant potential in food safety (Qi *et al.*, 2024).

Raman has excellent sensitivity and can indicate carotenoids through C–C and C=C strong spectral signatures of the conjugated in functional modes (Dhanani et al., 2022) (Mehta et al., 2021).Raman analysis can also monitor the vibrations of functional groups in eggs and can benchmark the quality of eggs (Kopec *et al.*, 2022).

Interpreting Raman spectral data presents numerous challenges, mainly stemming from its high-dimensional nature. A key obstacle is commonly known as the 'large p, small n' problem in statistics, where the number of variables (p) is substantial compared to the number of observations (n). This statistical challenge is exacerbated by the 'curse of dimensionality,' making direct modeling of high-dimensional spectral data prone to overfitting. To tackle this issue, two classical approaches, partial least squares (PLS) and principal components regression (PCR), have been introduced. Despite their utility, recent advancements in chemometrics, particularly in variable selection techniques, have significantly enhanced the predictive capacity, interpretability, and computational efficiency of multivariate calibration models (Zhu *et al.*, 2023b). Hence, exploring the application of Raman spectroscopy with variable selection methods for food analysis emerges as a promising research avenue.

Previous studies on egg freshness primarily utilized visible transmission spectroscopy on the eggshell (Karoui *et al.*, 2006; Kemps *et al.*, 2006; Liu *et al.*, 2007; Abdanan Mehdizadeh *et al.*, 2014; Dong *et al.*, 2019). Some methods only measure egg quality factors (De Ketelaere *et al.*, 2004; Cherian and Quezada, 2016; Zita, Jeníková and Härtlová, 2018), which will always have some errors and is time-consuming. It is noticeable that most of the studies conducted with Raman have a comparative aspect between eggs with different hatching conditions, including differences in the type of chicken and the type of nutrition (Cicek, 2009; Hesterberg *et al.*, 2012; Cherian and Quezada, 2016; Kopec *et al.*, 2022).

Considering the significant nutritional value of eggs and their pivotal role in the industry, ensuring their quality becomes imperative. In this article, Raman spectroscopy, known for its non-destructive nature, speed, and accuracy, has been used to explore the impact of temperature on egg yolk quality at 38 degrees for three hours. Raman spectra was taken at different intervals. Additionally, the influence of time on the yolk was investigated by keeping it at a constant 15 degrees for 48 hours, with Raman spectra taken 24 and 48 hours later. Biomarkers extracted from the Raman results further contribute to the analysis. The correlation between Raman peaks of the yolk and shell with total weight, yolk weight, egg index, yolk index, and yolk coefficient were examined to assess Raman's efficacy in measuring egg quality. The potential of Raman in evaluating egg quality through the shell was also assessed different time intervals. Finally, the PLS-DA statistical method was employed to classify Raman data based on storage time and temperature, aiming to discern differences between subgroups of classification.

## 2. Material and methods

*2.1. Eggs*

Twenty-six fresh eggs were divided into three groups of A, B, and C for analysis, then they were situated in different setups. Initially, their quality factors were measured and afterwards, their Raman spectra were taken. Group A included eight eggs that were kept at room temperature (38°C). At the inception, the egg quality measurements were determined, subsequently, Raman was taken from their shell (were named SHA), The eggs were carefully broken, the yolk quality measurements were determined and Raman was taken from them again(This subgroup called $At_0$) two hours after maintenance in the room constant condition, the Raman spectra were taken from them again (This subgroup called $At_1$), finally three hours after the start of the experiment, the Raman was taken from them once more (This subgroup called $At_2$). Group B included eight eggs that were situated in the (15°C). Identical to group A, first, all the egg quality measurements were determined afterward Raman was taken from their shell (were named SHB), eggs were carefully broken, the yolk quality measurements were defined and Raman was taken from them (This subgroup called $Bt_0$). egg yolks were kept at a constant temperature in the refrigerator for forty-eight hours. Further, the spectrum was taken from the yolk in the first twenty-four (This subgroup called $Bt_1$) hours and the second twenty-four hours (This subgroup called $Bt_2$). Group C contained ten chicken eggs and the basic measurements related to the shell (This subgroup called SHC) and yolk were incorporated as in the preceding two groups. It is noteworthy that the measurements were taken from the shell at intervals of four and three days.

*2.2. Destructive measurement techniques for egg quality*

Total weight, egg weight, egg shape index, yolk index, and yolk coefficient were calculated to measure the quality and freshness of the eggs. The Yolk coefficient is associated with the yolk weight and height Eq. (1), egg shape index Eq. (2) and yolk index Eq. (3) are associated with egg height, yolk height, and egg width, yolk width. Length, width, and height measurements were made with calipers and weight measurements were carried out with laboratory scales.

$$\text{Yolk coefficient} = \text{yolk weight} / \text{yolk height} \times 100 \qquad \text{Eq. (1)}$$

$$\text{Egg shape index} = \text{egg height} / \text{egg width} \times 100 \qquad \text{Eq. (2)}$$

$$\text{Yolk index} = \text{yolk height} / \text{yolk width} \times 100 \qquad \text{Eq. (3)}$$

The yolk index was measured by the method described by Paul Francis Sharp (Sharp and Powell, 1930). Yolk coefficient calculated by the Yande Liu method (Liu *et al.*, 2007). Egg shape index was determined using the Ebubekir Altuntasß method (Altuntaş and Şekeroğlu, 2008).

*2.3. Raman spectroscopy*

A micro-Raman instrument (Technooran, Microspectrophotometer-ram-532-004) was used, which has a laser with a wavelength of 532 nm and a maximum power of 200 mw (fig. 1) to acquire Raman spectra of egg yolk and eggshell with acquire time of 4s and laser power of 50 mw. To check the freshness and the quality of eggs the Raman spectra were taken in different conditions. To explore the impact of temperature on egg yolk quality at 38 degrees for three hours, Raman spectra was taken at two and three-hour intervals. Additionally, the influence of time on the yolk was investigated by keeping it at a constant 15 degrees for 48 hours, with Raman spectra taken 24 and 48 hours later.

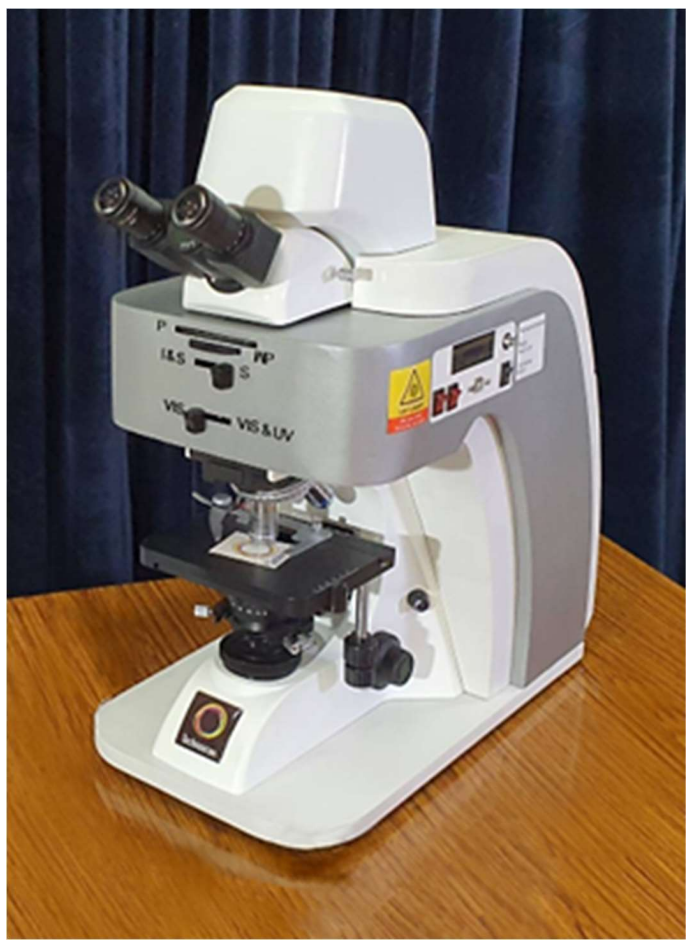

**Fig. 1**. Technooran, Microspectrophotometer-ram-532-004 instrument

*2.4. Data pre-processing*

One preprocessing method involves the averaging of spectra, a practice employed to decrease the quantity of wavelengths or to achieve spectrum smoothing. Another preprocessing technique is standardization (Cozzolino *et al.*, 2011).

The pre-processing of Raman spectra involved baseline correction and smoothing. Following normalization for boxplot, Raman peaks were computed using Python. To assess peak ratios, the ratios of two peaks were initially determined, and the average was subsequently calculated for each subgroup individually. Moreover, to elucidate the impact of time and examine the correlation between Raman peaks and egg qualities, instances within each subgroup were averaged, and their correlation was then computed. Graphs illustrating the Raman spectrum under various conditions were generated by averaging each subgroup. Notably, any missing values were handled using Python.

*2.5. Statistical analysis*

All Raman plots were epitomized by Origin (Origin pro-2016), and all statistical analyses were performed with Python (Jupiter notebook). Data analyzed by Tukey and one-way ANOVA test to statistical significance. p-value < 0.05 were considered statistically significant in one-way ANOVA test. Also, PLS-DA (Partial least square discriminant analysis) actuated as modeling of Raman data. In this model 840 feathers (Raman Intensity) used as X-data matrix and Y-matrix indicates the membership of each sample ($At_0$, $At_1$, $At_2$, $Bt_0$, $Bt_1$, $Bt_2$, C).

PLS-DA is a kind of statistical method that is supervised discriminant analysis. And can deal with classification and discriminant problems.(Xie *et al.*, 2023) PLS-DA classification performance is evaluated in terms of accuracy, specificity and sensitivity. Here, accuracy refers to the ability of a group to classify correctly. Accuracy and error rate are used to evaluate the best classification (Brasil, Cruz-Tirado and Barbin, 2022; Sahachairungrueng *et al.*, 2023).

### 3. Results and discussion

*3.1. Qualitative analysis of egg yolk and shell Raman Spectra*

Conducting Raman analysis enables the extraction of material information within each segment through the identification of Raman peaks. The Raman Spectroscopy results from egg yolk reveal the presence of eight distinct peaks spanning the range of 950 to 3000 $cm^{-1}$. The band that corresponds to the wavelength of 1004 $cm^{-1}$, is caused by the existence of protein (Kopec *et al.*, 2022). The next peak stands at 1156 $cm^{-1}$, and is caused by carotenoids (Liu, Chi and Chi, 2023). This bond is attributed to C–H in-plane bending and C–C stretching vibrations of the polyene chain(Mehta *et al.*, 2021). The wavenumber of 1301 $cm^{-1}$ induced by CH2 torsional states in lipids (Czamara *et al.*, 2015). the intensity of peaks 1451 $cm^{-1}$ indicates fatty acids and lipids(Kopec *et al.*, 2022). The stretching vibration of the C=C alkyl chain in egg yolk, which is present in carotenoids, causes a band at the wavenumber of 1518$cm^{-1}$ (Liu, Chi and Chi, 2023). 1667 band is related to C=C stretching vibrations in cholesterol. The band that exists in 2934 $cm^{-1}$ is correspond to the symmetrical stretching vibrations of the C–H in $CH_3$ moieties (Czamara *et al.*, 2015). Asymmetric stretching vibration of choline methyl leads to 3057 $cm^{-1}$(Bai *et al.*, 2020)( Fig. 2, Table. 1).

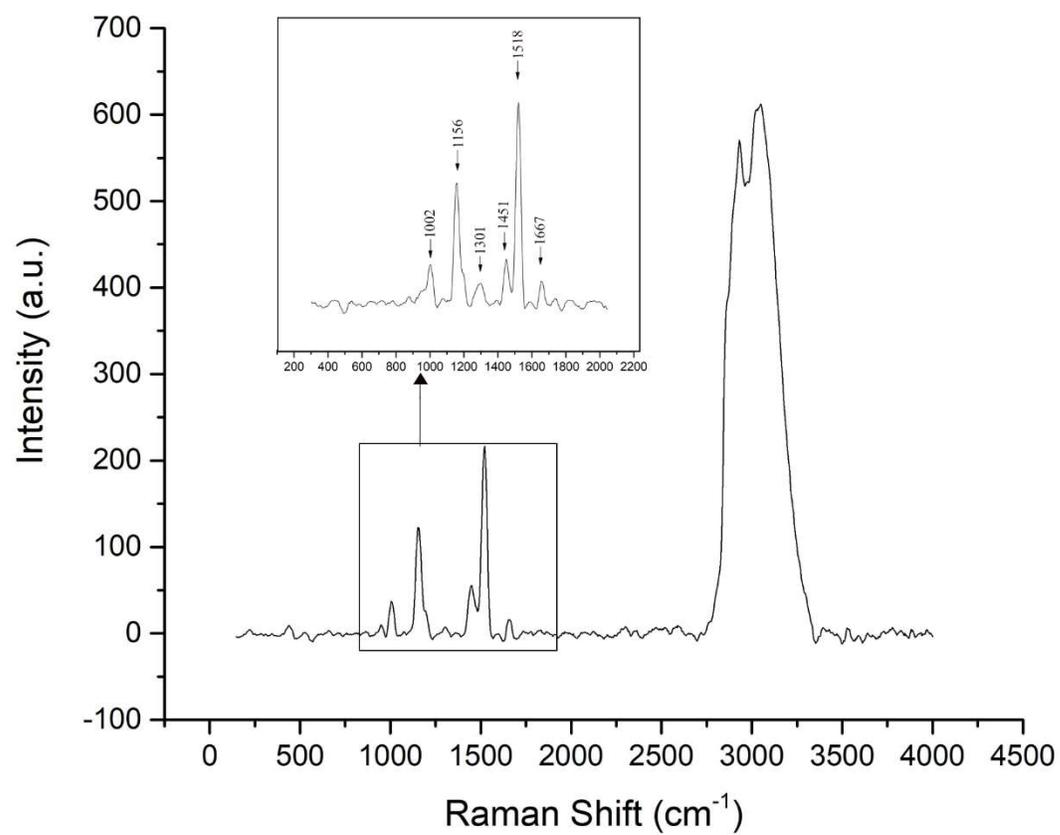

**Fig. 2.** Egg yolk Raman spectrum, 8 egg yolk peaks in 1002, 1156, 1301, 1451, 1518, 1667, 2934 and 3057 cm$^{-1}$

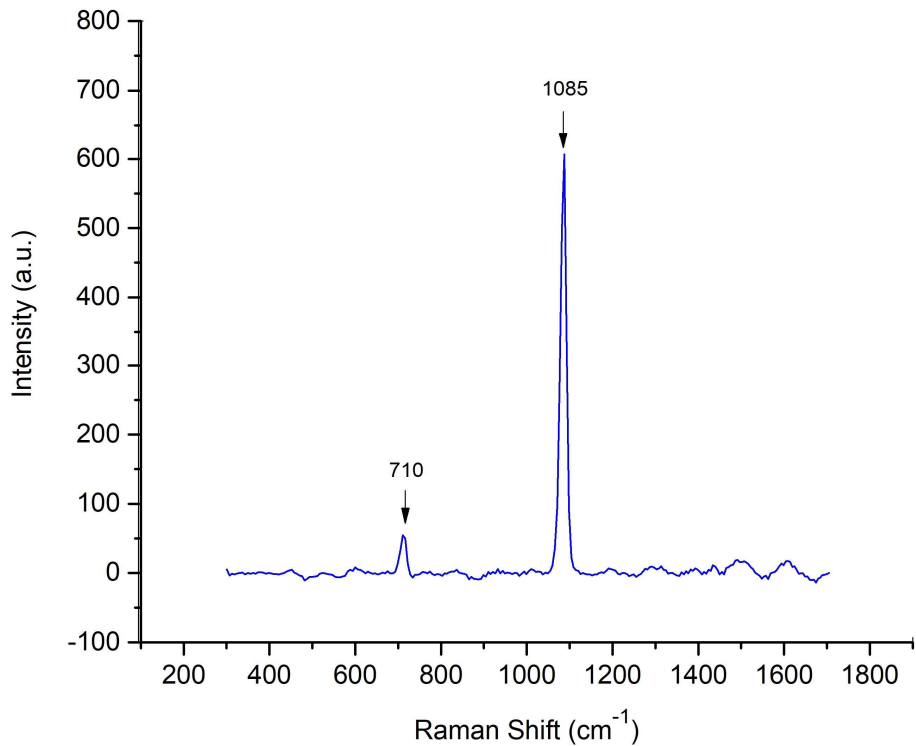

**Fig. 3**. Egg shell Raman spectrum, 2 packages of egg shells at 710 and 1085 cm$^{-1}$

The shell surface also has an exuberance of information, there was a layer of egg cuticles on the surface of the eggshell based on the research, which was the liquid protein that predominantly covered or embedded the pores on the surface of the eggshell. As a scattering spectrum, Raman spectroscopy can reflect the information on the egg shell surface (Siulapwa, Mwambungu and Mubbunu, 2014).

The Raman spectrum of an eggshell at an excitation wavelength of 532 nm was tested in this work, as shown in Fig. (3) it had two acute peaks, which were 710 cm$^{-1}$ and 1085 cm$^{-1}$ which indicate the presence of calcite ($CaCO_3$) (Thomas *et al.*, 2015)( Table. 1).

**Table. 1.** The vibrational assignment of the egg yolk and shell Raman spectra peaks

| Wavenumber [cm$^{-1}$] | Tentative assignments | |
|---|---|---|
| 1002 | protein | EYP1 |
| 1156 | carotenoids | EYP2 |
| 1301 | Lipids | EYP3 |
| 1451 | Lipids, fatty acid | EYP4 |
| 1518 | carotenoids | EYP5 |
| 1667 | C=C stretching vibrations | EYP6 |
| 2934 | C-H stretching vibrations | EYP7 |
| 3057 | choline methyl vibration | EYP8 |
| 710 | $CO_3^{2-}$ symmetric deformation | ESHP1 |
| 1085 | $CO_3^{2-}$ symmetric stretching | ESHP2 |

*3.2. Investigation of changes in temperature and storage time using egg yolk Raman spectra*

To assess the influence of storage conditions on egg quality, eggs in group A underwent exposure to a temperature of 38°C for three hours, and Raman measurements were conducted at the commencement of the study, two hours after the experiment initiation, and three hours thereafter. Similarly, eggs in group B were subjected to a consistent temperature of 15°C for 48 hours, with Raman readings taken at the experiment's outset, 24 hours into it, and finally, at the 48-hour mark. The Raman spectral results indicated a gradual decline in egg peaks over time for both groups, with some peaks becoming undetectable. Particularly noteworthy, the Raman graphs depict a significant reduction in the second, fifth, and seventh peaks associated with carotenoids.

Carotenoids, identified as the primary constituents influencing yolk color, play a crucial role in egg yolk composition (Hammeishøj, 2011). Factors such as temperature and time can lead to a decline in carotenoid content. Elevated temperatures induce nonenzymatic browning reactions, resulting in the loss of carotenoids, sugars, and amino acids, as well as the formation of Maillard reaction products (Carvalho et al., 2019). Beside the degradation of proteins in the egg yolk can decline the Raman peak. One of the factors that has played a prominent role in the breakdown of proteins during storage is related to proteases. It has also been suggested that degradation of egg yolk proteins during storage may affect the stability of lipids in egg yolk (Gao et al., 2017). Moreover, the abundance of yolk proteins, particularly those belonging to the lipocalin family, decreases with increasing temperature, alongside a rise in storage time (Gao *et al.*, 2017) Also, with increasing storage time, the oxidation of proteins and lipids increases (Tian, Lin and Bao, 2023).

Considering this information, it can be inferred that in group A, temperature exerts an influence, resulting in reduced Raman peaks, while in group B, the impact is attributed to the passage of time.

For a more in-depth analysis, peaks associated with the yolk were isolated and subjected to statistical testing using Tukey's tests (Fig. 5).

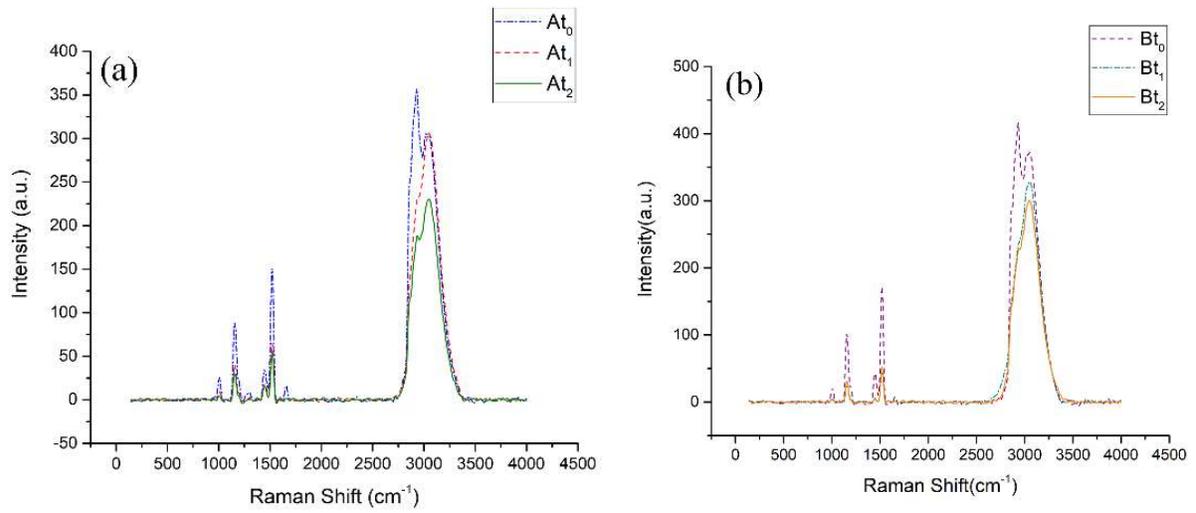

**Fig. 4.** Raman spectra of eggs in different temperature and storage time. a. Raman spectra related to group A, storing egg yolks at 38°C for 3 hours. b. Raman spectra related to group B, storing egg yolks at 15°C for 48 hours

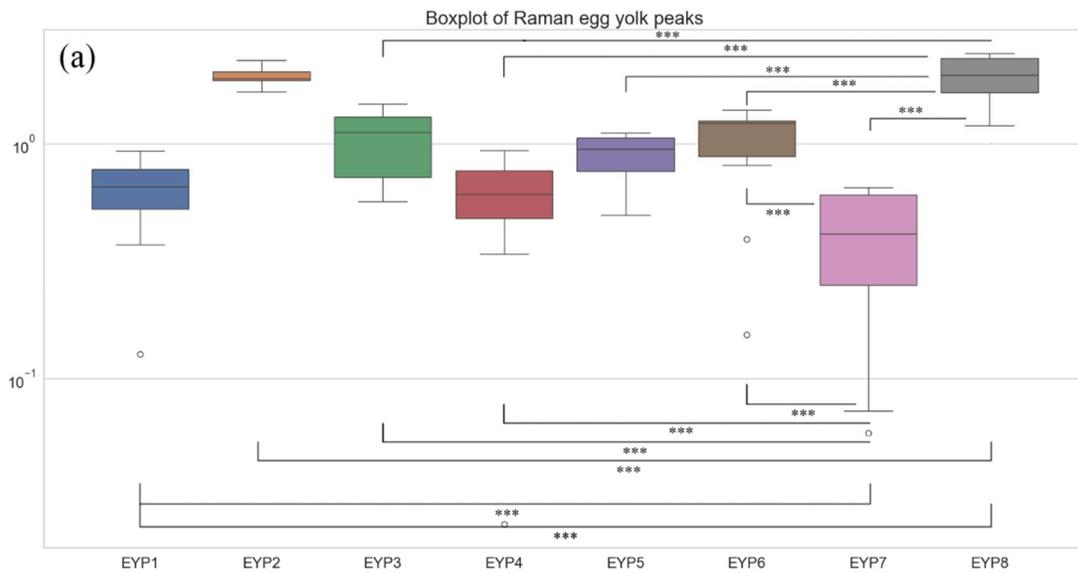

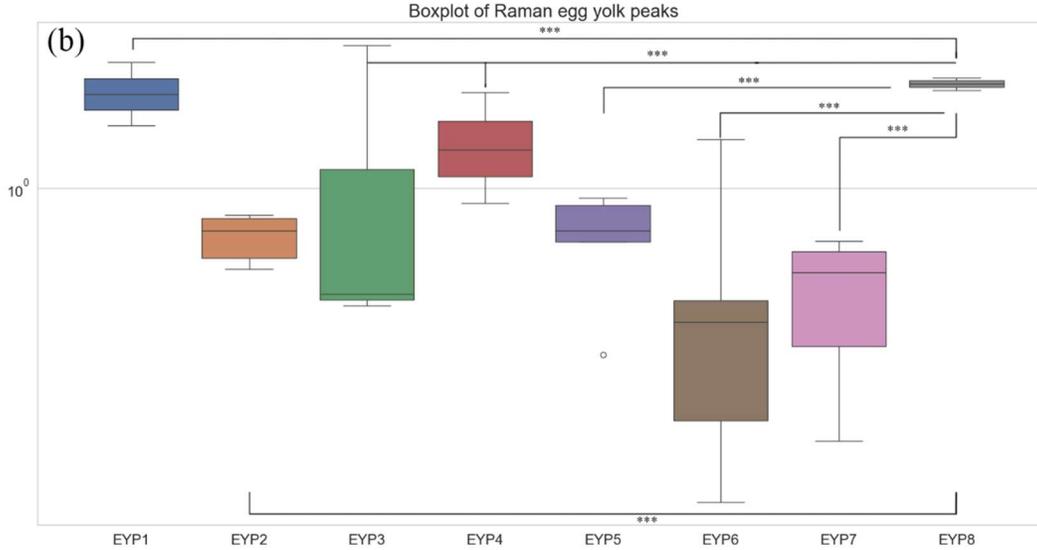

**Fig. 5.** Boxplot of Raman egg yolk peaks. a. Raman egg yolk peaks of group A. b. Raman egg yolk peaks of group B. (Raman peaks were obtained using Python and the statistically significant results have been marked with *** p-value< 0.05)

Based on the findings presented in Table. (1) and the frequency distribution depicted in Fig. (2), we have identified specific Raman bands, namely 1002, 1301, 1451, 3057, and 1156 cm-1, for comparative analysis. To discern the variations in intensity within each group, we opted to compute the following ratios for Group A: EYP1/EYP8, EYP1/EYP4, EYP3/EYP4, and EYP3/EYP8. Additionally, for Group B, we calculated the ratios EYP3/EYP8, EYP2/EYP4, EYP4/EYP3, and EYP2/EYP3. The resulting ratios are presented in Table. (2) and Fig. (5) for a comprehensive overview.

**Table .2.** Raman egg yolk peak intensity ratios of group A and B

| Ratio | $At_0$ | | $At_1$ | | $At_2$ | |
|---|---|---|---|---|---|---|
| | Mean | SD | Mean | SD | Mean | SD |
| EYP1/ EYP8 | 0.068 | 0.032 | 0.037 | 0.053 | 0.026 | 0.052 |
| EYP1/ EYP4 | 1.24 | 0.291 | 0.54 | 0.60 | 0.48 | 0.718 |
| EYP3/ EYP4 | 1.120 | 0.215 | 0.56 | 0.350 | 0.45 | 0.415 |
| EYP3/ EYP8 | 0.063 | 0.035 | 0.0401 | 0.029 | 0.025 | 0.0208 |
| Ratio | $Bt_0$ | | $Bt_1$ | | $Bt_2$ | |
| | Mean | SD | Mean | SD | Mean | SD |
| EYP3/ EYP8 | 0.042 | 0.029 | 0.021 | 0.014 | 0.016 | 0.029 |
| EYP2/ EYP4 | 4.62 | 5.6 | 2.81 | 2.2 | 1.58 | 2.36 |
| EYP4/ EYP3 | 5.35 | 4.38 | 2.75 | 2.61 | 1.38 | 2.37 |
| EYP2/EYP3 | 14.16 | 12.28 | 3.74 | 2.43 | 2.76 | 2.23 |

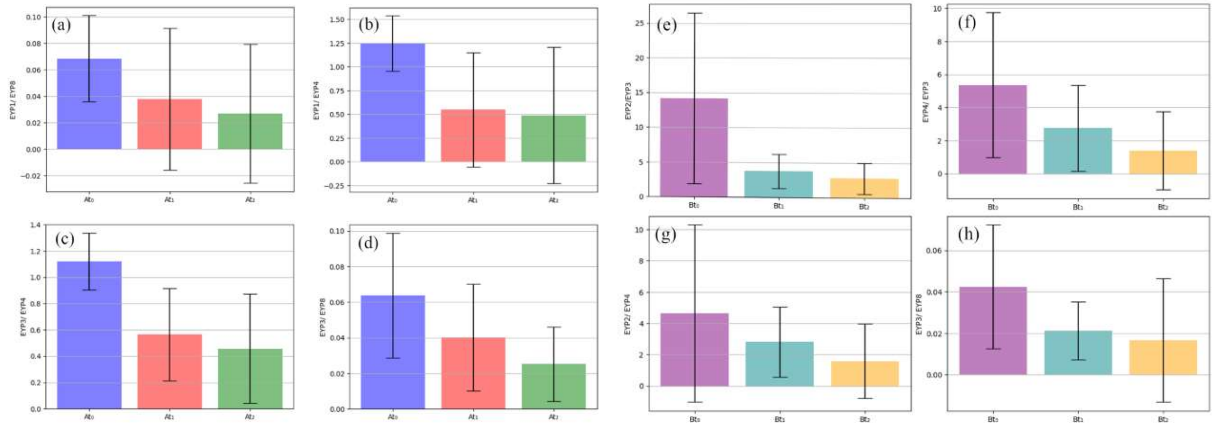

**Fig. 6.** Bar charts of the egg yolk peak ratios. From group A. a. EYP1/ EYP8. b. EYP1/ EYP4. c. EYP3/ EYP4. d. EYP3/ EYP8. from group B. e. EYP3/ EYP8. f. EYP2/ EYP4. g. EYP4/ EYP3. h. EYP2/EYP3

Examining Fig. (6) and Table. (2), it is evident that the presented ratios serve as valuable indicators for distinguishing differences between the groups. The analysis of vibration ratios associated with protein, choline methyl, lipids, fatty acids, and carotenoids reveals distinct variations for $At_0$ and $Bt_0$ within each group. Notably, the lowest ratios are consistently associated with $At_2$ and $Bt_2$, highlighting that fresh eggs exhibit higher proportions across all assessed parameters in both groups. Specifically, the ratios of lipids, fatty acids, and choline methyl in each group emerge as meaningful biomarkers for discerning the quality and freshness of yolks.

*3.3 Investigation of correlation between Raman spectra and egg quality characteristics*

To assess the correlation between Raman peaks and egg quality characteristics (Table. 3), we initially conducted Raman measurements on the eggshell. Subsequently, the quality components of each egg were quantified, followed by Raman analysis on the egg yolk. These sequential procedures were performed for each subgroup over time intervals of 4 and 3 days. The recorded characteristics of each subgroup were averaged, and a correlation graph was generated (see Fig. 7).

**Table. 3**. Statistical summary of egg quality characteristics

|       | total weight | egg weight | egg shape index | yolk index | yolk coefficient |
|-------|--------------|------------|-----------------|------------|------------------|
| Count | 26.000000    | 26.000000  | 26.000000       | 20.00000   | 22.000000        |
| Mean  | 57.541538    | 18.100385  | 132.236559      | 23.946730  | 66.058212        |
| Std   | 3.952996     | 1.369223   | 3.969495        | 3.796806   | 6.416632         |
| Min   | 51.150000    | 14.340000  | 121.824104      | 15.80906   | 59.371691        |
| 25%   | 54.945000    | 17.695000  | 130.352131      | 21.657906  | 63.166717        |
| 50%   | 56.910000    | 18.220000  | 132.388701      | 24.240856  | 64.774328        |
| 75%   | 60.095000    | 18.820000  | 135.025587      | 26.494078  | 67.087810        |
| max   | 67.350000    | 20.800000  | 138.251880      | 30.670217  | 87.211740        |

Examining the correlation plot (Fig. 7), a robust negative relationship (p-value < 0.5) is evident between the egg coefficient and PEY1, PEY4, and PESH1. Conversely, a strong positive correlation exists between the egg coefficient and PEY3, PEY5, and PEY6 (It's important to note that this study does not explore the relationship between egg quality factors). Furthermore, a pronounced positive correlation (p-value < 0.005) is identified between the yolk index and PESH2. The egg shape index (p-value < 0.5) displays a substantial negative correlation with PEY2, while showcasing a strong positive correlation with PEY8 and PEY7. Notably, a compelling positive correlation (p-value < 1) is observed between the yolk weight and PEY2. Lastly, it is noteworthy that a robust positive relationship (p-value < 0.5) is evident between total weight and PEY3, while a strong negative correlation exists between total weight and PESH1.

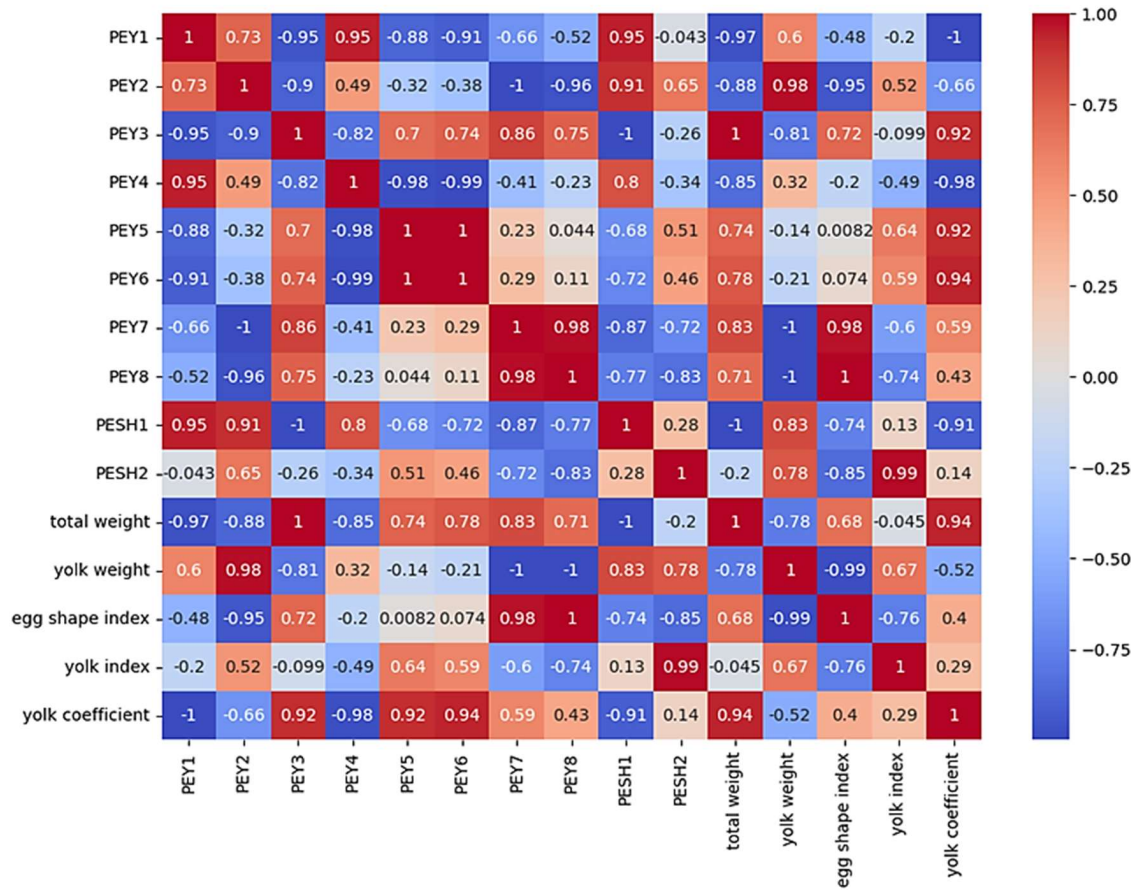

**Fig. 7**. Correlation plot between Raman peaks and egg quality characteristics

A negative correlation exists between the concentration of cholesterol and both yolk weight and egg weight, as indicated by Zita, Jeníková, and Härtlová in 2018(Zita, Jeníková and Härtlová, 2018). Analysis of peaks extracted from the yolk serum, which encompasses egg yolk cholesterol, reveals that nearly all of these peaks align with those found in the yolk itself, as noted by Hui-Jun et al. in 2012(Hui-Jun *et al.*, 2012). Consequently, the connection between some peaks and the overall weight, as well as the yolk weight, can be elucidated.

Moreover, the yolk coefficient exhibits a direct proportionality to its weight. Consequently, the association between peaks and the yolk coefficient is elucidated for this reason. Additionally, the yolk coefficient serves as an indicator of egg freshness and demonstrates a robust negative correlation with storage time, as highlighted by Abdanan Mehdizadeh et al. in 2014(Abdanan Mehdizadeh *et al.*, 2014). Hence, this observation can also be regarded as indicative of some other certain correlations.

According to the studies that have been done so far, a diet containing calcium improves the quality of the egg shell and improves the nutrients in it(Mikulski *et al.*, 2012). To some extent, this issue can explain the positive relationship between some yolk peaks and egg shell. yolk index determines egg freshness so it can explain the relationship between yolk index and peaks (Aryee et al., 2020).

*3.4. Investigating the effect of time on egg shell using Raman spectra*

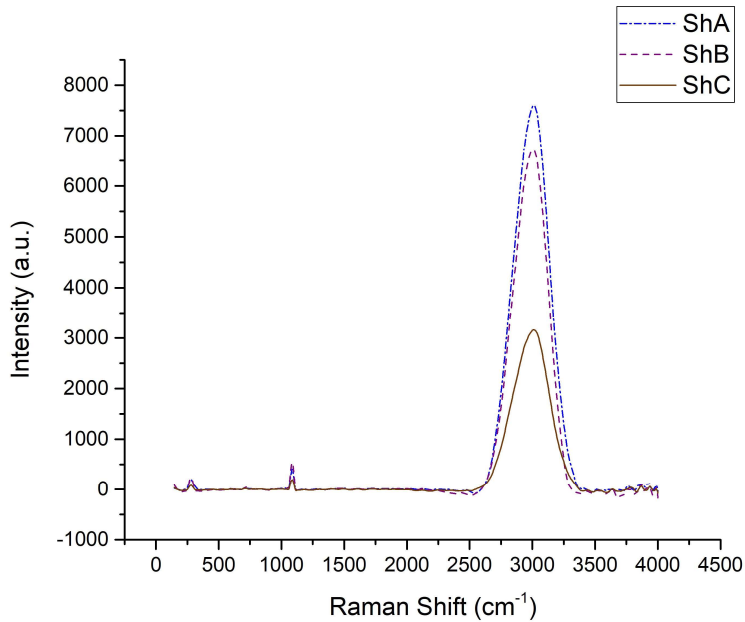

**Fig . 8**. Raman diagram of the eggshell at the time of starting the experiment (ShA), four days later (ShB) and three days later (ShC)

To showcase the effectiveness of Raman spectroscopy in assessing the freshness of eggs based on their shells, Raman spectra were captured from the eggshells within each group at distinct time intervals. The findings revealed a correlation wherein the peak intensity in the Raman diagrams exhibited a decline proportional to the duration of storage under consistent conditions.

During storage, moisture from egg is lost through evaporation at a rate that is influenced by temperature of the storage environment. Carbon dioxide is also lost through the shell pores while oxygen gets into the egg and creates an air bubble inside in-place of moisture and carbon dioxide, causing the egg to float when placed in water due to loss of weight (Eke, Olaitan and Ochefu, 2013). The movement of carbon dioxide and moisture through the egg shell increases the pH of the albumen and the yolk, decreases moisture percentage of egg albumen and decreases the albumen weight. There is a marked increase of naturally occurring psychophilic bacteria, coliform, staphylococci, yeast and moulds on egg shell surface and in egg content during egg storage (Eke, Olaitan and Ochefu, 2013). This may increase eggshell defects and thus lead to a decrease in eggshell $Ca_3Co_2$ peaks during storage. The term 'shell defects' encompasses the presence of any type of abnormality resulting from the distribution of calcium carbonate in the shell (Wengerska, Batkowska and Drabik, 2023) (see Fig 8).

*3.5. Classification of eggs at different temperature and storage time using egg yolk Raman spectra*

The primary objective of this study was to differentiate hen eggs based on their Raman analysis. In this section, Raman data specific to Group A (eggs subjected to a higher temperature for a shorter duration) and Group B (eggs exposed to a lower temperature for a longer duration) were collected. The exploratory Partial Least Squares Discriminant Analysis (PLS-DA) was conducted on a total of twenty-four samples, segregated into two sets (Fig. 9): a validation set and a calibration set, with a ratio of 20%. PLS-DA demonstrated a precise discriminating capability for both groups A and B.

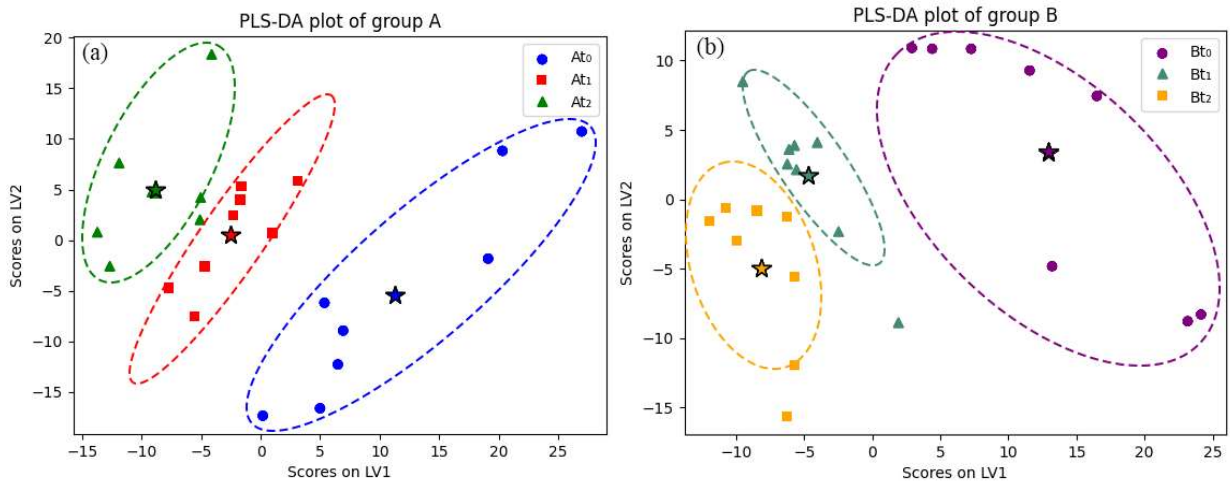

**Fig. 9.** PLS-DA plot of egg yolk Raman spectra different temperature and storage time. a. The chart related to group A, storing egg yolks at 38°C for 3 hours. b. The chart related to group B, storing egg yolks at 15°C for 48 hours

The overall correct classification rate was 80% in groups A and B. As can be seen in the Fig. (9), the eggs of subgroups $At_0$ and $Bt_0$, which had higher Raman peaks and their peak difference with other subgroups (especially in the range of 1000 to 2000 cm$^{-1}$), were well separated from subgroups $At_1$, $At_2$, and $Bt_1$, $Bt_2$. The accuracy of the model for classifying the data related to subgroups $At_0$ and $Bt_0$ was 100%. And the eggs of subgroups $At_1$, $At_2$, and $Bt_1$, $Bt_2$, their Raman was close to each other, the correct classification rates for each of them were 80. Regarding the first subgroups of each group, it is worth saying that all peaks somehow played a role in separating them from other subgroups, but perhaps the biggest contribution in separating them can be attributed to EYP2, EYP5, EYP7, and EYP8. Carotenoids and choline methyl play a significant role in subgroup separation. In the second and third subgroups ($At_1$, $At_2$, $Bt_1$, $Bt_2$) of each group, it is somehow not possible to attribute all the peaks for separation, and EYP2, EYP5, EYP8 have the greatest effect in separating them from other subgroups. It should be noted that the contribution of the eighth peak in separating these subgroups is greater than other peaks. In this subgroup, carotenoids and choline methyl have an important effect in separating these two subgroups from other subgroups of each group (Fig. 9, Table. 4).

Table. 4. Accuracy, sensitivity and specificity for group A and B

|  | Accuracy (%) | Sensitivity (%) | Specificity (%) | Error rate (%) |
|---|---|---|---|---|
| $At_0$ | 100 | 100 | 100 | 0 |
| $At_1$ | 80 | 100 | 75 | 20 |
| $At_2$ | 80 | 50 | 100 | 20 |
| $Bt_0$ | 100 | 100 | 100 | 0 |
| $Bt_1$ | 80 | 50 | 100 | 20 |
| $Bt_2$ | 80 | 100 | 60 | 33 |

## 4. Conclusions

Micro-Raman spectroscopy was employed to assess the quality and freshness of eggs. Dividing a set of 26 eggs into groups and subgroups, both shells and yolks underwent Raman analysis, revealing 8 peaks in the yolk and 2 peaks in the shell. The Raman results obtained from egg yolk and shell revealed 8 peaks in the yolk and two peaks in the shell. After exposing the yolks of 8 eggs to 38°C for three hours, a significant reduction in Raman peaks, particularly those associated with carotenoids, was observed. This reduction was attributed to the Maillard reaction. Similarly, maintaining the yolks at a constant 15°C for 48 hours led to reduced Raman peaks, primarily due to protein and fat oxidation, along with carotenoid reduction.

Analysis of relative peaks (EYP1/EYP8, EYP1/EYP4, EYP3/EYP4, EYP3/EYP8 for group A and EYP3/EYP8, EYP2/EYP4, EYP4/EYP3, EYP2/EYP3 for group B) identified lipid, fatty acids, and choline methyl ratios as potential biomarkers for assessing egg quality and freshness.

Correlation plot results highlighted negative correlations between egg coefficient and PEY1, PEY4, and PESH1, while strong positive correlations existed with PEY3, PEY5, and PEY6. Positive correlations were observed between yolk index and PESH2, egg shape index with PEY2, and strong positive correlations with PEY8 and PEY7. Yolk weight positively correlated with PEY2, and total weight showed a strong positive correlation with PEY3 but a strong negative correlation with PESH1. These correlations underscore Raman's effectiveness in detecting egg quality.

Examining Raman peaks from the shells of 26 eggs at 4 and 3 days revealed decreasing peaks over time, suggesting Raman's capability to detect egg freshness from the shell. PLS-DA statistical method results for classifying egg yolk Raman data based on temperature and time showed effective separation of fresher, higher-quality egg yolks from other subgroups, achieving 100% accuracy for the first subgroups and 80% accuracy for every group overall.

The results of this research showed that the Raman peaks related to carotenoids can specifically represent the freshness and quality of egg yolk, and it was also found that Raman has a unique ability to detect the freshness and quality of eggs. This subject can be used in the food industry to control the quality of this important food item, although the number of samples examined in this experiment was small, and for more accurate conclusions, more samples and experiments are needed.